\documentclass[hidelinks,12pt]{article}

\usepackage{arxiv}

\usepackage[utf8]{inputenc} 
\usepackage[T1]{fontenc}    
\usepackage{hyperref}       
\usepackage{url}            
\usepackage{booktabs}       
\usepackage{amsmath,amsfonts,amssymb,amsthm,wasysym}       
\usepackage{nicefrac}       
\usepackage{microtype}      
\usepackage{cleveref}       
\usepackage{graphicx}
\usepackage{natbib}
\usepackage{doi}

\usepackage{caption}
\DeclareMathOperator{\Tr}{Tr}
\numberwithin{equation}{section}
\setcitestyle{square}

\title{Axisymmetric pseudoplastic thin films in planar and spherical geometries}

\author{Chris Reese \\
	Department of Physics\\
	Lewis and Clark Community College\\
	Godfrey, IL   62035\\
	\texttt{ccreese@lc.edu}}

\renewcommand{\headeright}{Pseudoplastic thin films}
\renewcommand{\undertitle}{Pseudoplastic thin films}
\renewcommand{\shorttitle}{}

\hypersetup{
pdftitle={Pseudoplastic thin films},
pdfsubject={},
pdfauthor={Chris Reese},
pdfkeywords={}}

\begin{document}

\maketitle

\begin{abstract}

A simplified, pseudoplastic rheology characterized by constant viscosity plateaus above and below a transition strain rate is applied to axisymmetric, gravitationally driven spreading of a thin fluid film with constant volume flux source in planar and spherical geometries.    The model admits analytical solutions for flow velocity and volume flux.  Shear thinning influence on layer evolution is investigated via numerical simulation.  Isoviscous, asymptotic behaviors are recovered in small and large transition stress limits.  The effect of viscosity ratio on layer extent agrees with scaling arguments.  For intermediate transition stress, a flow behavior adjustment is observed consistent with heuristic arguments.  Planar and spherical geometry solutions are in agreement for sufficiently small polar angle.     

\end{abstract}

\section{Introduction}

Gravitational spreading of one fluid into another of differing density, where flow is predominantly unidirectional (e.g., horizontal or down-slope) constitutes a gravity current.  Thin film, viscous gravity currents have been a subject of interest since the seminal work of \citet{huppert82a,huppert82b} resulting in a significant analytical, numerical, and experimental literature addressing many aspects of layer dynamics, see e.g., \citep{simpson99,huppert06,ancey07,ungarish09}.
 
While first developed for isoviscous fluids, thin film theory has been extended to include many rheological behaviors.  For example, temperature dependent viscosity is of particular importance in geophysical contexts \citep{stasuik93,bercovici94,balmforth04,algwauish19}.  Also, strain rate dependent viscosities (i.e., generalized newtonian fluids) are ubiquitous in natural and engineering applications.  Viscoplastic rheologies such as Bingham and Herschel-Bulkley models find application to lava dome evolution \citep{blake90,balmforth99,balmforth00,balmforth06}. 
 Another behavior of practical interest is pseudoplasticity.  A pseudoplastic, or shear thinning, fluid is characterized by an effective viscosity which decreases with increasing strain rate.  This behavior is observed in polymeric \citep{bird87} and geophysical \citep{lavallee07,vasseur23} fluids.
Shear thinning, power law constitutive models admit various analytical approaches in the thin layer limit \citep{gratton99,perazzo03,myers05,nguetchue07}.  Other shear thinning rheologies include Sisko, Cross, and Carreau models.  Often, approximations to these models are adopted (e.g., \citep{wrobel20,james21}) in the interest of analytical simplicity (however, cf. \citet{pritchard15}).

Geometries addressed in the literature are primarily two-dimensional and 
axisymmetric flows on a plane.  However, thin film theory has also been extended to curved substrates \citep{oron97,roy02,kang16,kang17,taranets19}.  \citet{takagi10} considered flow and fingering instability of thin films on a cylinder and sphere with vertical gravitational acceleration.  Fluid spreading on a sphere with radial gravitational acceleration was considered in a geodynamical context \citep{reese10,reese11}.  This geometry may find application to planetary scale (i.e., low spherical harmonic degree) superplume head evolution (e.g., \citet{bercovici96,kerr04}) and/or global scale thermochemical diapir dynamics \citep{watters09} on terrestrial planets and icy satellites.   

In this work, a simplified pseudoplastic rheology \citep{james21} is adopted to investigate axisymmetric gravity currents in planar and spherical geometries.   The generalized newtonian viscosity (Sec 3.1) is a three parameter, piecewise linear function characterized by small and large strain rate viscosity plateaus and a rheological transition stress.  Section 2 is a heuristic, analysis of flow dynamics.  The model is reviewed in section 3.   Section 4 presents numerical results  addressing effects of pseudoplastic transition stress and viscosity ratio variation.  A summary and concluding comments are provided in section 5.  The appendix (section 6) outlines benchmarks of the numerical methods employed in the study.  

\section{Scaling analysis}
\label{sec:scaling}

Vertical and horizontal length scalings can be derived from dimensional analysis (e.g., \citet{griffiths93}) based on the fundamental, thin film caveat that layer thickness $H$ is small with respect to lateral extent $L$.  To first order in the aspect ratio $H/L$, scaling analysis implies: flow is primarily tangential to the surface, normal stress is negligible resulting in a hydrostatic pressure distribution, and the tangential pressure gradient is balanced by the shear stress gradient perpendicular to the surface.      

\subsection{Small strain rate}

In the isoviscous, small strain rate limit, $\dot{\varepsilon} \ll \dot{\varepsilon_c}$, spreading is controlled by viscosity $\eta$ (see Sec.\ \ref{sec:rheo}).  For lateral velocity scale  $V$, continuity implies that the vertical velocity is first order small in the aspect ratio $\dfrac{H}{L} \, V$.  The pressure scale $p \sim \rho g H$.  The pressure gradient is balanced by vertical shear stress gradient,
\begin{equation}
\frac{\rho g H}{L} \sim \frac{\tau}{H} \; ,
\end{equation}
where the shear stress scale $\tau \sim \eta \dfrac{V}{H}$ implying
\begin{equation}
\frac{\rho g H}{L} \sim \eta \frac{V}{H^2} \; .
\end{equation}
Kinematics require that the lateral velocity scale 
\begin{equation}
V \sim L/t \;, 
\end{equation}
where $t$ is time.  For a constant volume flux $Q$, mass conservation requires
\begin{equation}
Q t \sim L^2 H \; .
\end{equation}
Eliminating $H$ and $V$ between Eqs.\ (2.2,2.3,2.4) yields the lateral length scaling as a function of time
\begin{equation}
L \sim \left(\frac{Q^3 \rho g}{\eta} \right)^{1/8} \; t^{1/2} \, . 
\end{equation}
It follows that the layer height scales as
\begin{equation}
H \sim \left( \frac{\eta \, Q}{\rho g} \right)^{1/4} \; .
\end{equation}
 
\subsection{Large strain rate}
 
For large strain rate, $\varepsilon \gg \varepsilon_c$, spreading is controlled by the high strain rate viscosity $\mu$ (see Sec.\ \ref{sec:rheo}).  In this limit, the lateral and height scales are expected to 
be 
\begin{equation}
L_{\mu} \sim \chi^{1/8} \; L \,, \qquad \qquad H_{\mu} \sim \chi^{-1/4} \; H \,,
\label{eq:scale_chi}
\end{equation}
where the viscosity ratio $\chi = \eta/\mu$.  For intermediate strain rate, i.e.\ $\tau \sim \tau_c$, there is no asymptotic scaling for $L$ and $H$ throughout layer evolution as illustrated by numerical results described below.

\section{Model}

Consider a fluid of density $\rho$ and strain rate dependent generalized newtonian viscosity $\eta_{\rm eff}(\dot{\varepsilon})$ spreading on a rigid surface.  For characteristic velocity and length scales $V$ and $L$ respectively, a Reynolds number can be defined as the ratio of the momentum diffusion time to advection time,
\begin{equation}
 {\rm Re} = \frac{L^2/\nu}{L/V} = \frac{\rho V L}{\eta_{\rm eff}} \; ,
\end{equation}
where $\nu = \eta_{\rm eff}/\rho$ is the kinematic viscosity.      
Sufficiently small Re guarantees non-inertial flow reducing the Cauchy momentum equation
\begin{equation}
 - \nabla p + \nabla \cdot \boldsymbol{\tau} + \rho \, \boldsymbol{g} = 0 \; ,
\end{equation}
where $p$ is pressure and $\boldsymbol{g}$ is gravitational acceleration.    

For an incompressible fluid, the continuity equation 
\begin{equation}
\nabla \cdot \boldsymbol{u} = 0 \; .
\end{equation}
Free surface boundary conditions are zero pressure $p = 0$, zero traction 
$\boldsymbol{\tau} \cdot \hat{\boldsymbol{n}} = 0$, and the kinematic condition
specifying that a fluid parcel on the boundary remain on the boundary. 
Basal boundary conditions are no-slip (i.e., zero velocity component tangential to the basal surface), and matching of fluid velocity perpendicular to the layer base with
source velocity $w_s$.

\subsection {Pseudoplastic constitutive relation}  
\label{sec:rheo}

For a generalized newtonian rheology, the effective viscosity is a function of strain rate.  The deviatoric stress tensor
\begin{equation}
\boldsymbol{\tau} = 2 \, \eta_{\rm eff}(\dot{\varepsilon}) \, \boldsymbol{\dot{\varepsilon}} \; ,
\end{equation}
where
\begin{equation}
\boldsymbol{\dot{\varepsilon}} = \frac{1}{2} \left[ \nabla \boldsymbol{u} + \nabla \boldsymbol{u}^{\rm T} \right] \; ,
\end{equation} 
is the strain rate tensor and  
\begin{equation}
\dot{\varepsilon}  =  \left[ \frac{1}{2}  \; \Tr \left( \boldsymbol{\dot{\varepsilon}} \boldsymbol{\dot{\varepsilon}}^{\rm T} \right) \right]^{1/2} \,, \qquad
\tau   =  \left[ \frac{1}{2}  \; \Tr \left( \boldsymbol{\tau} \boldsymbol{\tau}^{\rm T} \right) \right]^{1/2} \; ,
\end{equation} 
are the strain rate and stress invariants, respectively. 

\begin{figure}[ht!]
\centering
\includegraphics{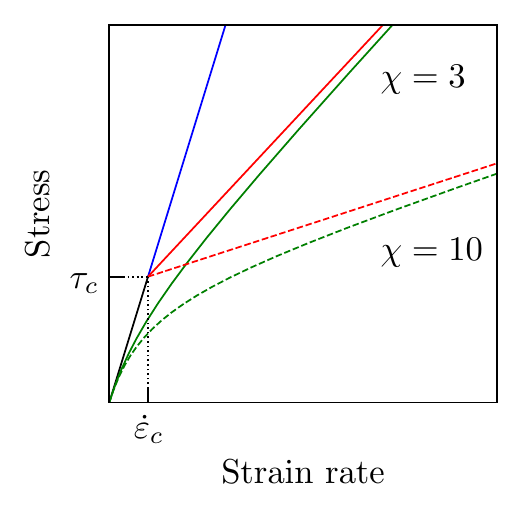}
\hspace{0.25in}
\includegraphics{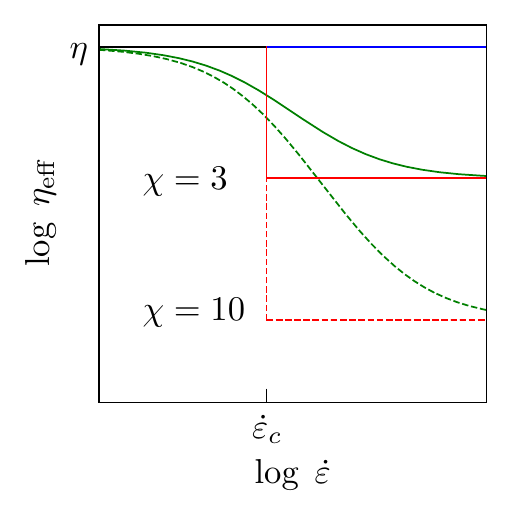}
\caption{A qualitative comparison of various rheologies.  (left) The stress invariant versus strain invariant for isoviscous (blue), simplified pseudoplastic (red), and Cross fluids (green).  All fluids have the same small strain rate viscosity.  The simplified pseduoplastic and Cross fluids approach the same asymptotic large strain rate viscosity. The parameter $\chi$ is the ratio of small strain rate to large strain rate viscosities, $\chi = \eta/\mu$.  Solid and dashed curves correspond to $\xi$ = 3 and 10, respectively.  (right) Effective viscosities of the rheological models.}
\label{fig1} 
\end{figure}

The piecewise linear approximation to a pseudoplastic rheology adopted in this study is
\begin{equation}
\eta_{\rm eff}(\dot{\varepsilon}) = \begin{cases}
\eta  & \dot{\varepsilon}  \le   \dot{\varepsilon}_{\rm c} \\
 \dfrac{\tau_c}{2  \dot{\varepsilon} } \left(1 - \dfrac{\mu}{\eta} \right) + \mu &  \dot{\varepsilon}  >   \dot{\varepsilon}_{\rm c} \\
\end{cases}
\label{eq:rheo}
\end{equation}
where $\dot{\varepsilon}_{\rm c}$ is the critical strain rate for onset of shear thinning,  $\tau_c = 2\, \eta \, \dot{\varepsilon}_{\rm c}$  is the critical stress invariant, $\eta$ is the low strain rate viscosity, and $\mu$ is the high strain rate viscosity.  This rheological model is compared to the isoviscous case and a Cross fluid in Fig.\ \ref{fig1}.

\subsection{Planar geometry}

In this section, the governing equations are non-dimensionalized and expanded, to leading order, in the thin layer limit for the rheological model described above.  This constitutive relation admits determination of the radial velocity profile and radial volume flux.  Vertical integration of the continuity equation yields the layer height evolution equation.  For cylindrical coordinates ($r$, $z$, $\phi$), the axisymmetric velocity 
\begin{equation}
\boldsymbol{u} = u(r,z,t) \, \hat{\mathbf r} + w(r,z,t) \, \hat{\mathbf z} \; .
\end{equation}  
The vertical locations of the flow base and free surface are $z=0$ and $z=h(r,t)$, respectively.
The gravitational acceleration $\boldsymbol{g} = - \, g \hat{\boldsymbol z}$.    

\subsubsection{Non-dimensionalization}
\label{sec:plan_nondim}

Motivated by heuristic arguments (Sec.\ 2), the governing equations are non-dimensionalized and expanded in the small layer aspect ratio limit.  For a rigorous formulation of the expansion in planar axisymmertry the reader is referred to \citet{balmforth00}.  
As in Section 2, let $H$ be the characteristic layer thickness and $L$ be the radial length scale.  The vertical coordinate $z$ and layer thickness $h$ are scaled by $H$ while pressure is scaled by $\rho g H$.  The radial coordinate $r$ is scaled by $L$. Continuity implies that the vertical velocity scale is first order small with respect to radial velocity.  To leading order, the radial velocity scale $V = \rho g H^3/\eta L$ representing a balance between lateral pressure gradient and vertical shear stress gradient.  Time is scaled by $L/V$.   In the thin layer limit, the aspect ratio $H/L$ is the small parameter in which the governing equations are expanded.    

\subsubsection{Pseudoplastic transition height}

To leading order, normal stress is negligible and gravitational body force is balanced by hydrostatic pressure subject to the free surface boundary condition, $\left. p \right|_{z=h} = 0$,
\begin{equation}
p = h \, \left(1 - \frac{z}{h} \right) \; .
\end{equation}
The radial pressure gradient due to layer height variation is balanced by vertical shear of the radial flow
\begin{equation}
\frac{\partial p}{\partial r} = \frac{\partial \tau_{rz}}{\partial z} \; .  
\end{equation}
Substituting for $p$, integrating, and applying the boundary condition $\left. \tau_{rz} \right|_{z=h} = 0$ yields  
\begin{equation}
\tau_{rz} = - h h^{\prime} \, \left(1-\frac{z}{h} \right) \; ,
\label{eq:tau} 
\end{equation}
where $h^{\prime} = \dfrac{\partial h}{\partial r}$.  The stress invariant $\tau = h |h^{\prime}| \, \left(1-\dfrac{z}{h} \right)$ increases as $z$ decreases from $z=h$ and there is a vertical level $Y$ where $\tau = \tau_c$,
\begin{equation}
Y  =  h \left[ 1 - \frac{\tau_c}{h |h^{\prime}|} \right] = h \left[ 1 - \frac{2 \, \dot{\varepsilon}_c}{h |h^{\prime}|} \right] \; .
\label{eq:Y}
\end{equation}

\subsubsection{Radial velocity}

The constitutive equation implies
\begin{equation}
\tau_{rz} =  2 \, \eta_{\rm eff}(\dot{\varepsilon}) \, \dot{\varepsilon}_{rz} = \eta_{\rm eff} \, (\dot{\varepsilon}) \dfrac{\partial u}{\partial z} \; ,
\end{equation}
where the strain rate invariant  $\dot{\varepsilon} = \dfrac{1}{2} \, \left| \dfrac{\partial u}{\partial z} \right|$. Let $u_-(z)$ be the radial velocity for $z<Y$.  Substituting for $\eta_{\rm eff}$, with radial velocity increasing monotically with height $\left(\dfrac{\partial u_-}{\partial z} > 0 \right)$ yields
\begin{equation}
\frac{\partial u_-}{\partial z} = - \chi h h^{\prime} \left(1 - \frac{z}{h} \right) - 2 \, (\chi-1) \, \dot{\varepsilon}_c \; ,
\label{eq:du_m}
\end{equation}
where $\chi = \dfrac{\eta}{\mu}$ is the low strain rate to high strain rate viscosity ratio.
Integrating with respect to $z$, and applying the no slip boundary condition 
$\left. u_- \right|_{z=0} = 0$,
\begin{equation} 
u_-(z) = - \chi h^2 h^{\prime} f\left(\frac{z}{h}\right) - 2 \, (\chi-1) \, h \, \dot{\varepsilon}_c \frac{z}{h} \;, \qquad 0 \le z \le Y \; ,
\end{equation} 
with $f(x) = x - x^2/2$. 

\begin{figure}[ht!]
\centering
\includegraphics{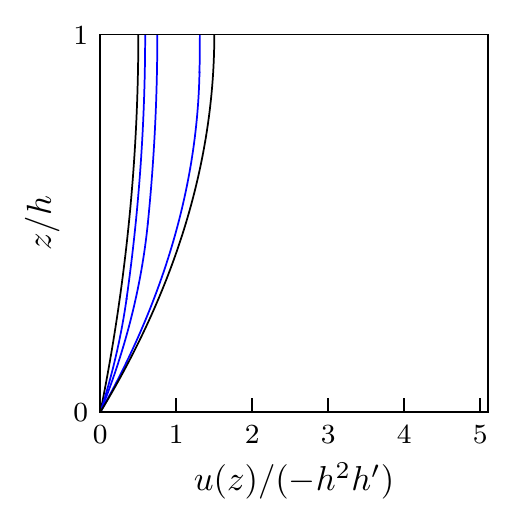}
\hspace{0.25in}
\includegraphics{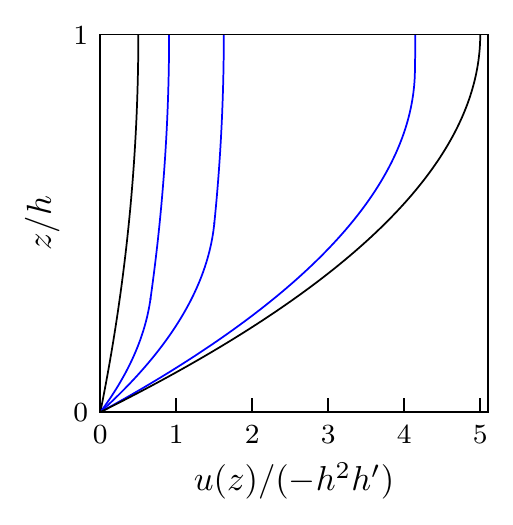}
\caption{Radial velocity profiles for the cases $\chi=3$ (left), $\chi = 10$ (right), and different pseduoplastic transition levels (left to right) $Y/h$ = (0, 0.3, 0.5, 0.9, 1).  The black curves correspond to the low strain rate and high strain rate isoviscous limits (see discussion in text).} 
\label{fig2}
\end{figure}

Above $z=Y$, $\eta_{\rm eff} = \eta$, and the momentum balance reduces to the isoviscous 
case.  Letting $u_+(z)$ be the velocity for $z>Y$,  
\begin{equation}
\frac{\partial u_+}{\partial z}  = - h h^{\prime} \left(1 - \frac{z}{h} \right)  \; .
\label{eq:du_p}
\end{equation} 
Integrating and matching velocities $u_-(Y) = u_+(Y)$ gives,
\begin{equation}
u_+(z) = - h^2 h^{\prime} f\left(\frac{z}{h}\right) - (\chi-1) \, h^2 h^{\prime} f\left(\frac{Y}{h}\right)-  2 \, (\chi-1) \, h \, \dot{\varepsilon}_c \frac{Y}{h} \;, \qquad Y \le z \le h \; .
\end{equation} 
Eliminating the transition strain rate using Eq. (\ref{eq:Y}),
\begin{equation}
u(z) = \begin{cases}
-h^2 h^{\prime} \left[\chi f\left(\dfrac{z}{h}\right) - (\chi-1) \left( 1 - \dfrac{Y}{h} \right) \dfrac{z}{h} \right] & 0 \le z \le Y \\[1em]
-h^2 h^{\prime} \left[ f\left(\dfrac{z}{h}\right) +(\chi -1) \dfrac{Y^2}{2 h^2}\right] & Y \le z \le h 
\label{eq:u}
\end{cases}
\end{equation} 
This velocity distribution is shown in Fig.\ \ref{fig2}.
In the limit $Y \rightarrow 0$, the velocity $u(z) = -h^2 h^{\prime} f(z/h)$ which corresponds to the $\eta_{\rm eff} \rightarrow \eta$ isoviscous limit.  Likewise, when $Y \rightarrow h$, $u(z) = -\chi h^2 h^{\prime} f(z/h)$ which is the $\eta_{\rm eff} \rightarrow \mu$ isoviscous limit.   

The height evolution equation (see next section) depends on the vertically integrated radial velocity, i.e., the radial volume flux per unit azimuthal length
\begin{equation}
U = \int_0^Y u_-(z) \, dz + \int_Y^h u_+(z) \, dz \; .
\end{equation}
Evaluating,
\begin{equation}
U = -h^3 h^{\prime} \left[\frac{1}{3}+ \frac{1}{2} \, (\chi-1) \left(\frac{Y}{h}\right)^2 \left(1 - \frac{Y}{3h}\right)\right] \; .
\label{eq:U}
\end{equation}
In the low strain rate $Y \rightarrow 0$ limit, $U$ reduces to the isoviscous  $\eta_{\rm eff} \rightarrow \eta$ case, i.e., $U = -h^3 h^{\prime} / 3$.  Likewise for $Y \rightarrow h$, $U = - \chi h^3 h^{\prime} / 3$ which is the $\eta_{\rm eff} \rightarrow \mu$ isoviscous limit.

\subsubsection{Evolution equation}

The free surface kinematic condition
\begin{equation}
\frac{\partial h}{\partial t} + u(r,h,t) \frac{\partial h}{\partial r} - w(r,h,t) = 0 \; .
\end{equation}
Vertically integrating the continuity equation over the layer height subject to basal and free surface boundary conditions and identifying $w(r,0,t) = w_s(r,t)$ yields the layer height evolution equation representing mass conservation,
\begin{equation}
\frac{\partial h}{\partial t} + \frac{1}{r} \frac{\partial}{\partial r} \left( r U \right)  = w_s(r,t) \; ,
\label{eq:plan_evol}
\end{equation}
where $U$ is given by Eq.\ (\ref{eq:U}).

\subsection{Spherical geometry}

The following section is a brief outline the thin layer expansion in spherical geometry.  It is shown that, to leading order, local flow on a sufficiently large sphere is insensitive to substrate curvature resulting in polar velocity and volume flux identical to the planar case. 
In spherical coordinates ($\theta$, $r$, $\phi$), the axisymmetric velocity field 
\begin{equation}
\boldsymbol{u} = u(\theta,r,t) \, \hat{\boldsymbol \theta} + w(\theta,r,t) \, \hat{\boldsymbol r} \; .
\end{equation}
The radial locations of the flow base and free surface are $r=R$ and $r=R+h(\theta,t)$, respectively.
The gravitational acceleration $\boldsymbol{g} = - \, g \, \hat{\boldsymbol r}$.      
The layer is considered to extend laterally along an arclength $R \, \theta_f(t)$.  The thin layer approximation $h \ll R \, \theta_f$ is assumed to hold throughout spreading given sufficiently large substrate curvature $R$.  Because $\theta_f \sim 1$, it follows that $h \ll R$. 

The radius of curvature $R$ provides an intrinsic lengthscale for non-dimensionlization.  Lengths are scaled by $R$ and pressure by $\rho g R$.  The polar  velocity scale $V = \rho g R^2/\eta$ and timescale is $R/V$.  Upon non-dimensionalizing, the quantity in which the governing equations are expanded is $h \ll 1$.
A change of variables is introduced 
\begin{equation}
r = 1 + \xi \; ,
\end{equation}   
where $0 \le \xi \le h$ is the non-dimensional radial coordinate measured from the spherical surface.      

The radial pressure gradient balances the gravitational body force subject to the free surface boundary condition $\left. p \right|_{\xi=h} = 0$,
\begin{equation}
p = h \, \left(1 - \frac{\xi}{h} \right) \; .
\end{equation}
The polar pressure gradient due to layer height variation is balanced by radial shear of the polar flow.  In terms of $\xi$, 
\begin{equation}
\frac{1}{(1+\xi)} \frac{\partial p}{\partial \theta} = \frac{1}{(1+\xi)^2} \frac{\partial}{\partial \xi}\left( (1+\xi)^2 \tau_{\xi \theta} \right)) \; .
\end{equation}
Substituting for $p$, applying the boundary condition $\left. \tau_{\xi \theta} \right|_{\xi=0}=0$, and dropping terms ${\cal{O}}(h^2)$ and higher,
\begin{equation}
\tau_{\xi \theta} = - h h^{\prime} \left( 1 - \frac{\xi}{h} \right) \; ,
\end{equation}
where $h^{\prime} = \dfrac{\partial h}{\partial \theta}$.  This linear stress distribution is identical to the planar case Eq.\ (\ref{eq:tau}).  Thus, the radial level $\xi=Y$ below which the stress invariant exceeds $\tau_c$ is given by Eq.\ (\ref{eq:Y}). 

To leading order, the constitutive equation
\begin{equation}
\tau_{\xi \theta} =  2 \, \eta_{\rm eff}(\dot{\varepsilon}) \, \dot{\varepsilon}_{\xi \theta} = \eta_{\rm eff} \, (\dot{\varepsilon}) \dfrac{\partial u}{\partial \theta} \; ,
\end{equation}
where the strain rate invariant  $\dot{\varepsilon} = \dfrac{1}{2} \, \left| \dfrac{\partial u}{\partial \theta} \right|$.   Analysis proceeds as in the planar case.   The polar velocities $u_{\pm}(\xi)$ above and below $Y$ satisfy Eqs.\ (\ref{eq:du_p}) and (\ref{eq:du_m}), respectively.  The polar velocity profile $u(\xi)$ and polar volume flux per unit azimuthal length $U$ are given by Eqs.\ (\ref{eq:u}) and (\ref{eq:U}), respectively, with $h^{\prime}  \rightarrow \dfrac{\partial h}{\partial \theta}$ and $z \rightarrow \xi$.  
  
To leading order, in terms of $(\theta,\xi,t)$, the free surface kinematic condition, 
\begin{equation}
\frac{\partial h}{\partial t} + u(\theta,h,t) \frac{\partial h}{\partial \theta} - w(\theta,h,t) = 0 \; .
\end{equation}
Radially integrating the continuity equation subject to basal and free surface boundary conditions, identifying $w(\theta,0,t) = w_s(\theta,t)$ and dropping terms $\mathcal{O}(h^2)$ and higher yields
\begin{equation}
\frac{\partial h}{\partial t} +  \left[ \frac{1}{\sin \theta} \frac{\partial}{\partial \theta} \left( \sin \theta \, U \right) \right]= w_s(\theta,t)  \; ,
\label{eq:sphr_evol}
\end{equation}
where $U$ is given by Eq.\ (\ref{eq:U})

\section{Numerical results}

\subsection{Planar geometry}

Variation of rheological transition stress and viscosity ratio are investigated numerically.  The transition stress range is chosen to include anticipated end-member behaviors.  The non-dimensional source function 
\begin{equation}
w_s(r) = w_0 \, (r_0^2 - r^2) \, H(r_0 - r) \; ,
\label{eq:plan_source}
\end{equation}
where $r_0$ = 0.15, $w_0$ = 0.1, and $H$ is the unit step function.   

\subsubsection{Height field}

\begin{figure}[ht!]
\centering
\includegraphics{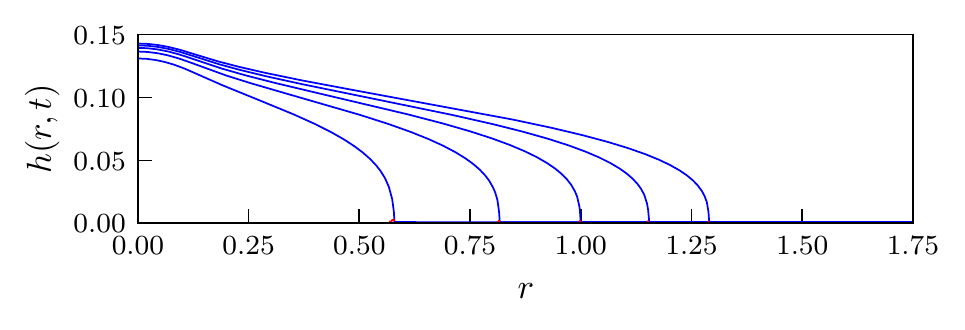}

\vspace*{-0.125in}
\includegraphics{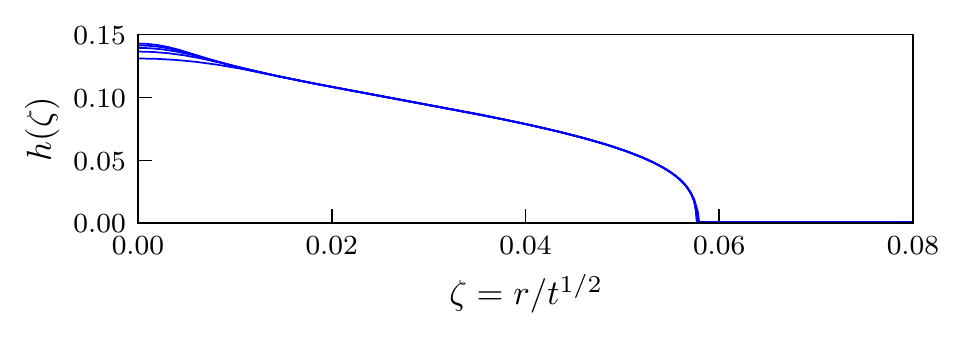}
\caption{Evolution for $\tau_c = 300 \times 10^{-4}$ and $\chi = 10$.  (top) Layer height field $h(r,t)$ (blue) together with the transition surface $Y(r,t)$ (red, dashed) for non-dimensional times (1, 2, 3, 4, 5) $\times$ 10$^3$.  The transition surface location $Y=0$ except for small regions around the flow front.  (bottom) Layer height field versus the similarity variable $\zeta = r/t^{1/2}$.  Away from the source, solution convergence onto the similarity form indicates that the fluid behaves isoviscously with $\eta_{\rm eff} \approx \eta$.} 
\label{fig:plan300}
\end{figure}

The effect of varying the pseudoplastic transition stress is considered for viscosity ratio $\chi$ = 10.  For large transition stress ($\tau_c = 300 \times 10^{-4}$), the location of the rheological transition surface $Y$ is indistinguishable from zero except for small regions near the flow front (Fig.\ \ref{fig:plan300}).  Also, the solution converges to the similarity solution in the similarity variable $\zeta = r/t^{1/2}$.  away from the source.  Thus, the fluid behaves isoviscously with $\eta_{\rm eff} \approx \eta$. 

\begin{figure}[ht!]
\centering
\includegraphics{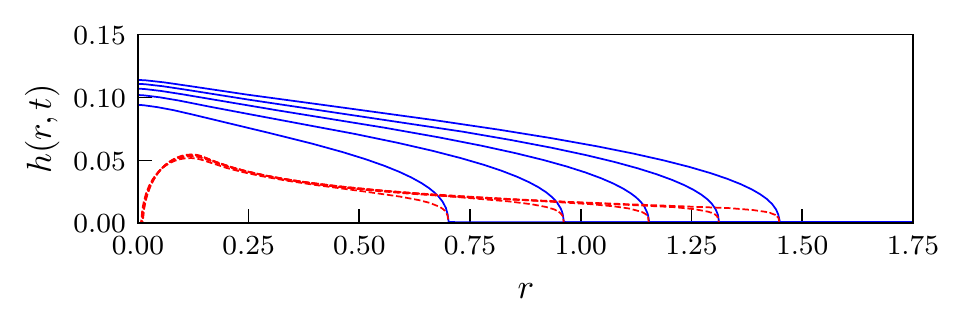}

\vspace*{-0.125in}
\includegraphics{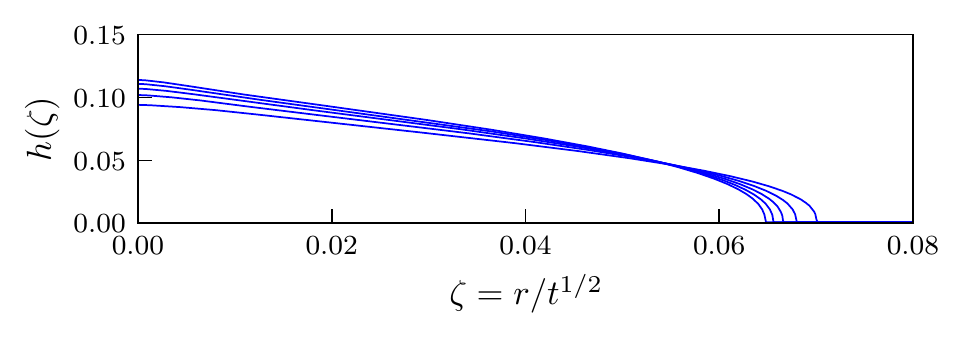}
\caption{As Fig.\ \ref{fig:plan300} with $\tau_c = 30 \times 10^{-4}$.  (top) Height field and pseudoplastic transition surface located at approximately half the layer height.  (bottom) Height field as a function of the similarity variable $\zeta$.  For intermediate transition stress, a similarity solution of the isoviscous form is not admitted.} 
\label{fig:plan030}
\end{figure}

For intermediate transition stress ($\tau_c = 30 \times 10^{-4}$), the location of the transition surface is initially approximately half of the layer height but decreases as evolution proceeds (Fig.\ \ref{fig:plan030}).  In this case, it is not possible to collapse the solution onto a similarity form.     

\begin{figure}[ht!]
\centering
\includegraphics{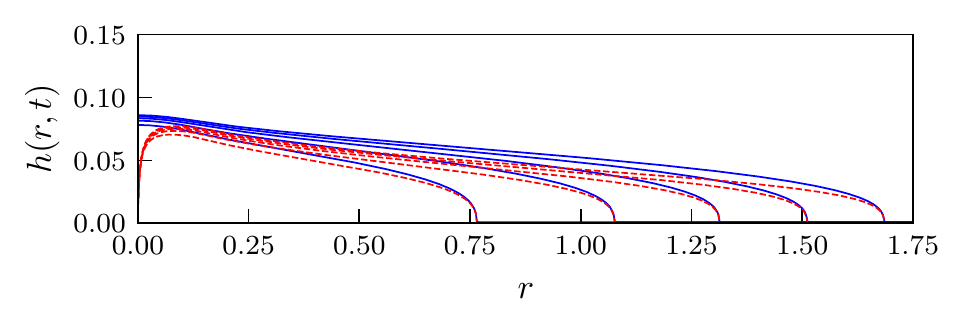}

\vspace*{-0.125in}
\includegraphics{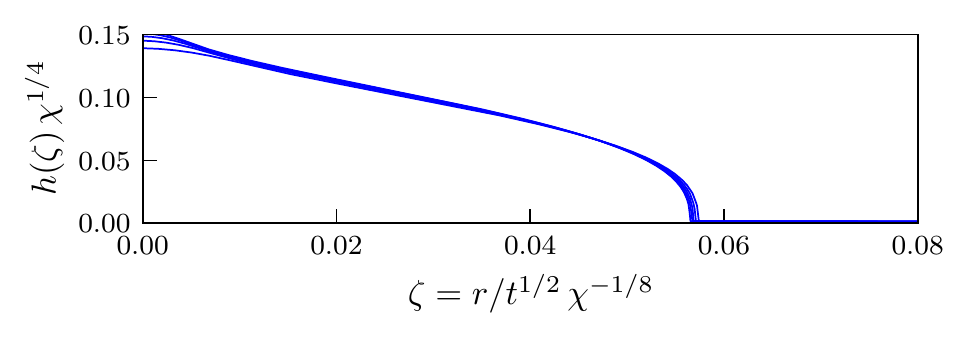}
\caption{As Fig.\ \ref{fig:plan300} with $\tau_c = 3 \times 10^{-4}$.   (top) Height field and pseudoplastic transition surface with location approximately equal to the layer height.  (bottom) Height field scaled by the viscosity ratio factor $\chi^{1/4}$ as a function of the similarity variable $\zeta$ scaled by the viscosity ratio factor $\chi^{-1/8}$.  The approximate convergence of the solution onto the similarity form suggests the flow is effectively isoviscous with $\eta_{\rm eff} \rightarrow \mu$.} 
\label{fig:plan003}
\end{figure}

For small transition stress ($\tau_c = 3 \times 10^{-4}$), the transition surface location is approximately equal to layer height (Fig.\ \ref{fig:plan003}).  Scaling layer height and radius by the appropriate viscosity ratio factors (Sec.\ \ref{sec:scaling}) indicates that the solution is converging to an isoviscous similarity solution with high strain rate viscosity $\eta_{\rm eff} \approx \mu$.      

\subsubsection{Flow front}

\begin{figure}[ht!]
\centering
\includegraphics{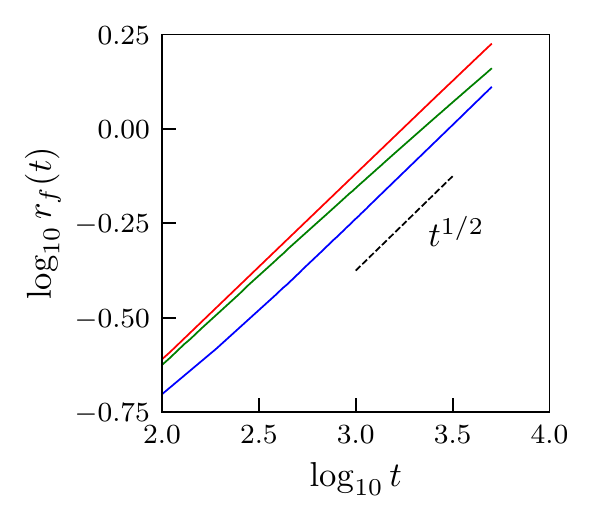}
\hspace{0.25in}
\includegraphics{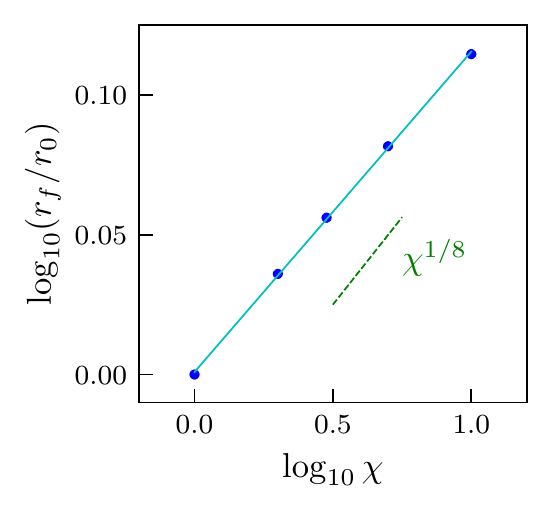}
\caption{(left) Variation in flow front radius with time for $\chi$=10 and pseudoplastic transition stress $\tau_c = (300, 30, 3) \times 10^{-4}$ (blue, greed, red). The black dashed line shows the characteristic, isoviscous scaling. (right) Flow front location after fixed evolution time $T = 5 \times 10^3$ for $\tau_c = 3 \times 10^{-4}$ as a function of viscosity ratio $\chi$. Results are in good agreement with the expected asympotic scaling (green dashed line).}   
\label{fig:plan_front}
\end{figure}

In Fig.\ \ref{fig:plan_front}, flow front location as a function of time is shown for viscosity ratio $\chi$ = 10 and three values of transition stress.  The flow front is defined by $h(r_f,t)$ = 0.01.  For sufficiently large $\tau_c$, flow is controlled by the low strain rate viscosity, i.e., $\eta_{\rm eff} \approx \eta$ and flow radius exhibits the asymptotic, isoviscous, time scaling.  Also, for small transition stress, the flow behaves isoviscously with $\eta_{\rm eff} \approx \mu$ throughout the evolution time considered.  In the case of intermediate $\tau_c$, flow radius behavior exhibits a transition.  Initially, layer evolution is consistent with the $\eta_{\rm eff} \approx \mu$ regime.  As the layer spreads, the stress invariant decreases, and the behavior approaches that for the small strain rate limit.

Scaling analysis (Sec.\ \ref{sec:scaling}) suggests that, for spreading initially controlled by $\mu$, shear stress decreases with time like $t^{-1/2}$.  That is, the dimensional stress invariant,
\begin{equation}
\tau \sim \left(\rho^3 g^3 \mu^5 Q \right)^{1/8} \; t^{-1/2} \; .
\end{equation}
A flow transition time scale $t_c$ can be defined as the time when $\tau \sim \tau_c$
\begin{equation}
t_c \sim  \frac{ \left(\rho^3 g^3 \mu^5 Q \right)^{1/4}}{\tau_c^2}    \; .        
\end{equation}
Non-dimensionalizing (Sec.\ \ref{sec:plan_nondim}),
\begin{equation}
t_c \sim \frac{Q^{1/4}}{\chi^{5/4} \tau_c^2} \; .
\end{equation}
For non-dimensional source function Eq.\ (\ref{eq:plan_source}), $Q = \dfrac{\pi}{2} w_0 r_0^4$
\begin{equation}
t_c \sim 10^3 \, \left(\frac{10}{\chi} \right)^{5/4} \, \left( \frac{30 \times 10^{-4}}{\tau_c} \right)^2 \; .
\end{equation} 
Thus, the intermediate transition stress case is expected to undergo this flow transition during the non-dimensional evolution time of the calculation $T = 5 \times 10^3$.  The low transition stress case considered would not exhibit such behavior until a non-dimensional time $t \sim 10^2 \, T$.

For fixed transition stress and evolution time, increasing viscosity ratio $\chi$ implies increasing flow radius (Eq.\ \ref{eq:scale_chi}).  For $\tau_c = 3 \times 10^{-4}$ and $T = 5 \times 10^3$, flow radius was calculated as a function of $\chi$.  Results agree with the scaling analysis (Fig.\ \ref{fig:plan_front}).         

\subsection{Spherical geometry}

In spherical geometry, the non-dimensional source function 
\begin{equation}
w_s(\theta) = w_0 \, (\theta_0^2 - \theta^2) \, H(\theta_0 - \theta)
\end{equation}
with $\theta_0$ = 0.15 and $w_0$ = 0.1.  This source corresponds to non-dimensional volume flux $Q$ identical to the planar geometry volume flux to $\mathcal{O}(\theta_0^5)$.  In the following sections, the effects of varying rheological transition stress and viscosity ratio are investigated numerically.    

\subsubsection{Height field}

\begin{figure}[ht!]
\centering
\includegraphics{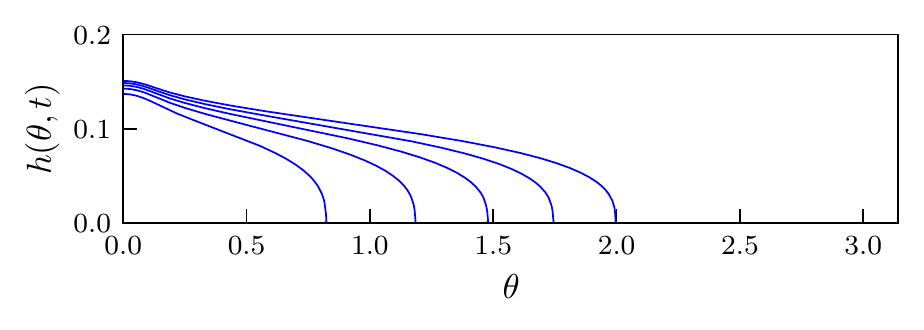}

\vspace*{-0.125in}
\includegraphics{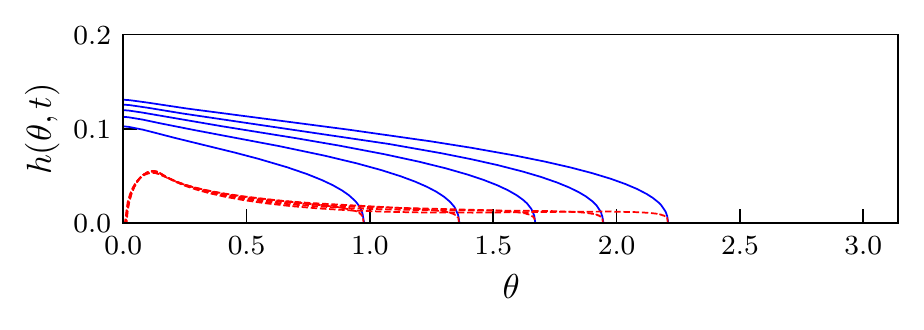}

\vspace*{-0.125in}
\includegraphics{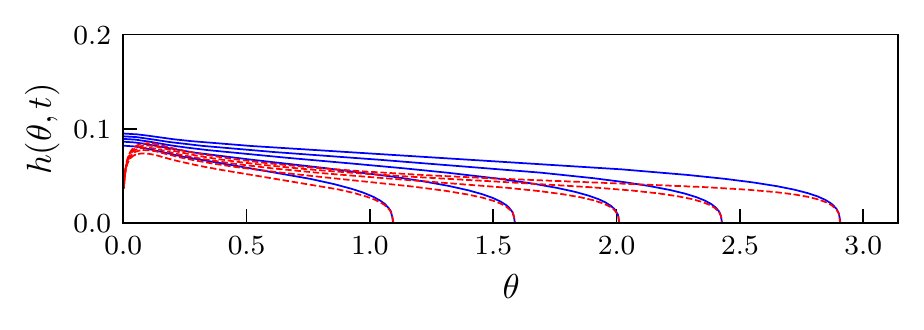}
\caption{Evolution in spherical geometry for $\chi = 10$ and, from top to bottom, $\tau_c = (300, 30, 3) \times 10^{-4}$, respectively.  Layer height fields $h(\theta,t)$ (blue) together with the transition surface $Y(\theta,t)$ (red, dashed) for non-dimensional times (2, 4, 6, 8, 10) $\times$ 10$^3$.  (top) The transition surface location is indistinguishable from $Y=0$.  (middle) The low viscosity region constitutes a decreasing fraction of the layer height throughout evolution.  (bottom) The high viscosity part of the flow is only a small fraction near the free surface.} 
\label{fig:sphr_height}
\end{figure}

The range of rheological transition stresses investigated in spherical geometry is the same as that for the planar geometry case.  Likewise, the viscosity ratio $\chi$ = 10.  Results are summarized in Fig.\ (\ref{fig:sphr_height}).  For large transition stress ($\tau_c = 300 \times 10^{-4}$), the location of the rheological transition surface $Y \sim 0$.  In this limit, the fluid behaves isoviscously with $\eta_{\rm eff} \approx \eta$.  For intermediate transition stress ($\tau_c = 30 \times 10^{-4}$), the low viscosity part of the flow is initially approximately half of the layer height.  The low viscosity layer height fraction decreases as evolution proceeds.  For small transition stress ($\tau_c = 3 \times 10^{-4}$), $Y \sim h$.                 

\subsubsection{Flow front}

\begin{figure}[ht!]
\centering
\includegraphics{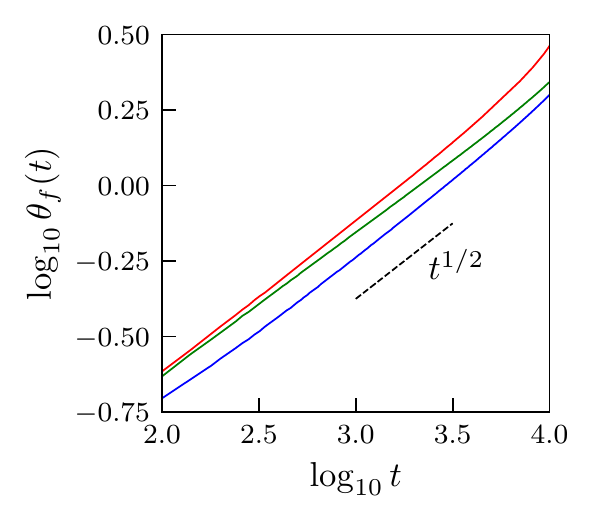}
\includegraphics{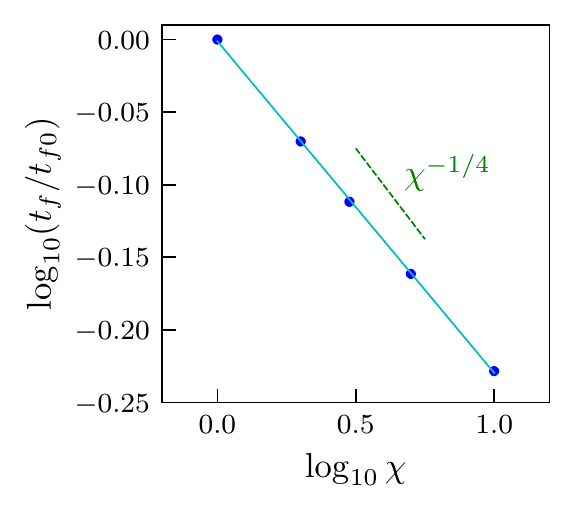}
\caption{(left) Variation in flow front polar angle with time for $\chi$=10 and pseudoplastic transition stress $\tau_c = (300, 30, 3) \times 10^{-4}$ (blue, greed, red). The black dashed line shows the characteristic, isoviscous scaling.  (right) Time for spreading to equatorial polar angle as a function of viscosity ratio $\chi$.  Flow behavior is in agreement with asymptotic scaling analysis (green dashed line). }
\label{fig:sphr_front}
\end{figure}

The flow front location is defined by $h(\theta_f,t)$ = 0.01.  As in the planar case, the flow front location is controlled by high (low) viscosities for large (small) rheological transition stress (Fig.\ \ref{eq:sphr_front}).  In the intermediate $\tau_c$ case, flow dynamics transition during evolution due to decreasing low viscosity layer fraction.  
In spherical geometry, isolation of the viscosity ratio effect is complicated by transition to a converging flow front when spreading proceeds past the equatorial polar angle.  To isolate rheological influence on layer dynamics, the time for spreading to $\theta_f \sim \pi/2$ is calculated as a function of viscosity ratio $\chi$.  Scaling analysis (Sec.\ \ref{sec:scaling}) suggests that the time for spreading to polar angle $\theta_f$ for $\eta_{\rm eff} \sim \eta$,
\begin{equation}
t^{\ast} \sim \left( \frac{Q^3 \rho g}{\eta} \right) \theta_f^2 \; .
\end{equation}
In the asymptotic limit, $\eta_{\rm eff} \sim \mu$,
\begin{equation}
t_{\mu}^{\ast} \sim \chi^{-1/4} \; t^{\ast} \; .
\end{equation}
Numerical results (Fig.\ \ref{fig:sphr_front}) are in good agreement with the anticipated scaling. 
\begin{figure}[h!]
\centering
\begin{minipage}{0.65 \textwidth}
\includegraphics{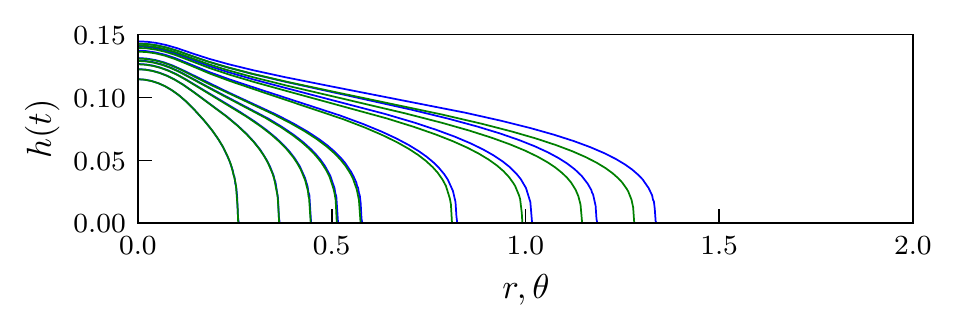}
\end{minipage}%
\begin{minipage}{0.35 \textwidth}
\includegraphics{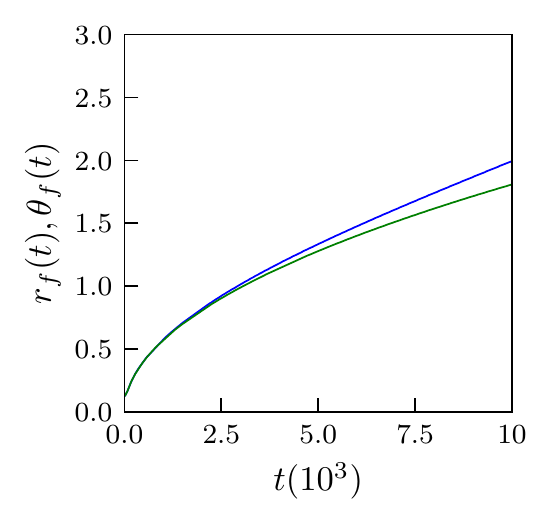}
\end{minipage}
\begin{minipage}{0.65 \textwidth}
\includegraphics{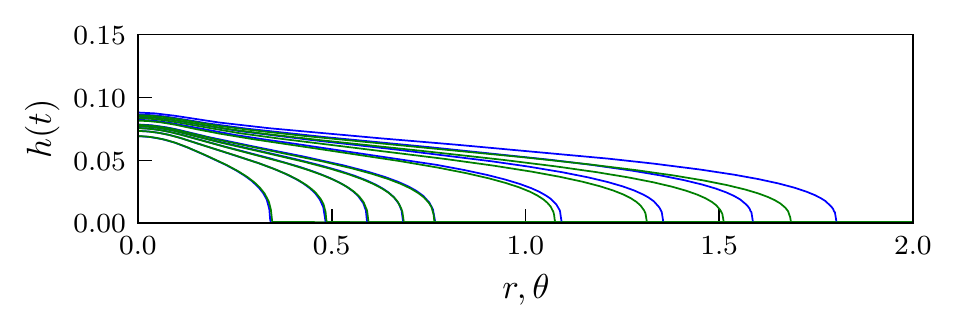}
\end{minipage}%
\begin{minipage}{0.35 \textwidth}
\includegraphics{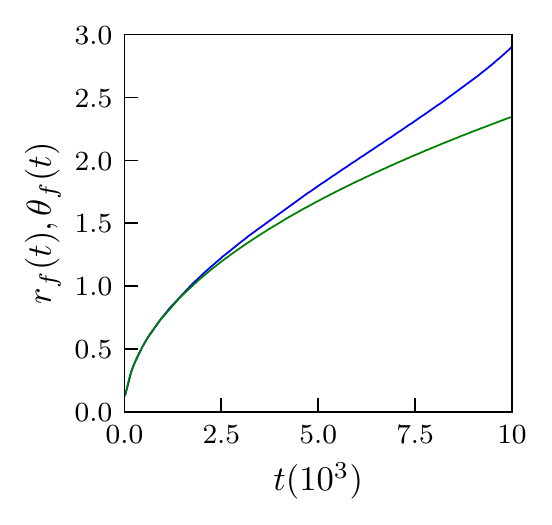}
\end{minipage}
\caption{(top,left) Height fields $h(\theta,t)$ and $h(r,t)$ in spherical (blue) and planar (green) geometries for the isoviscous (or large transition stress) and non-dimensional times (0.2, 0.4, 0.6, 0.8, 1, 2, 3, 4, 5) $\times$ 10$^3$.  (top,right) Flow front location as a function of time.  For polar angle $\theta \lesssim 0.5$, the solutions are in good agreement. (bottom) As for the top figures for the case $\tau_c = 3 \times 10^{-4}$ and $\chi$ = 10. }
\label{fig:sphr_plan}
\end{figure}

\subsection{Small polar angle approximation} 

In the small polar angle limit $\theta \ll 1$, the non-dimensional evolution equation for layer height in spherical geometry reduces to that for planar geometry.  Planar and spherical solutions for identical source functions should be equal in the small polar angle approximation.  Layer height fields and flow front location for the two geometries are compared in Fig.\ (\ref{fig:sphr_plan}) for the isoviscous and small transition stress cases.  Solutions are in agreement for polar angle $\theta \lesssim 0.5$.  For example, the relative difference in flow front location at this angle is $\sim$ 0.3 \%

\section{Conclusions}
A pseudoplastic rheology was applied to axisymmetric thin film evolution in planar and spherical geometries with constant volume flux source.  Closed form expressions for velocity profile and volume flux were derived.  The numerical approach utilized in the study was benchmarked against previous analytical and numerical solutions.  Influence of rheological transition strain rate and viscosity ratio on layer evolution was explored numerically.  In the limits of large and small transition strain rate, approximately isoviscous evolution was observed.  For intermediate transition strain rate, control of layer spreading can undergo an adjustment from high to low strain rate viscosity.  Planar and spherical geometry solutions agree for sufficiently small polar angle (see e.g., \citet{takagi10}). 

While admittedly simplified, the rheological model captures bulk features of shear thinning and allows for efficient exploration of parameter space.  It may find applicability in geodynamical contexts including, but not limited to, plume head evolution \cite{bercovici96}, isostatic adjustment to thermochemical diapirs \citep{watters09}, and glacial dynamics \citep{schoof10}.  The rheology is also adaptable to channelized flow \citep{sochi15} and, as silicic magmas exhibit shear thinning behavior \citep{jones20,vasseur23}, could be implemented in models of dike emplacement and/or magma flow in volcanic conduits \citep{gonnermann07}.
The pseudoplastic model is readily extendable to spreading on substrates with constant radius of curvature and vertical gravitational acceleration \citep{takagi10} relevant to industrial coating applications.  Finally, modification of the model to more complex, non-planar surfaces \citep{lin12,lin21} is also a possibility.       
                 
\section{Appendix: Numerical method}

In this appendix, the numerical method is benchmarked against analytical solutions and other numerical approaches.  Layer height evolution equations (Eqs.\ \ref{eq:plan_evol},\ref{eq:sphr_evol}) are of the form
\begin{equation}
\frac{\partial h(r,t)}{\partial t} = \mathcal{D}\left[h(r,t)\right]
\end{equation}
where $\mathcal{D}$ is a non-linear differential operator.  The Python package py-pde \citep{zwicker20} provides methods for solving partial differential equations of this form.  The py-pde, method-of-lines scheme utilizes implicit, Adams backward differentiation \citep{hindmarsh83,petzold83}. 
To benchmark the numerical approaches, the isoviscous, constant volume flux source case is considered and compared to analytical and numerical solutions. 
  
\subsection{Planar geometry}

\begin{figure}[ht!]
\centering
\begin{minipage}{0.65 \textwidth}
\includegraphics{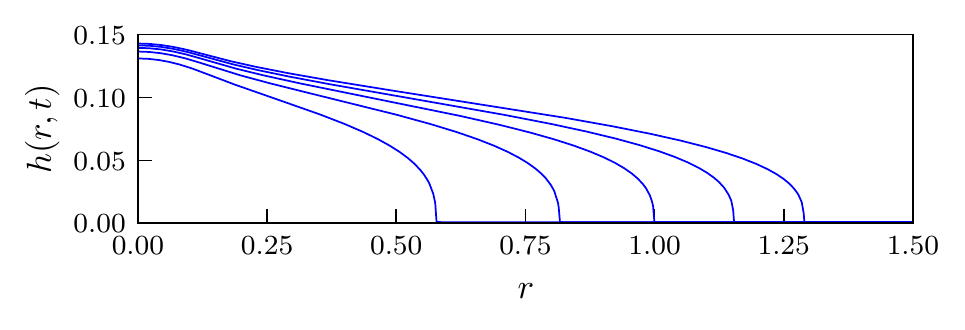}

\includegraphics{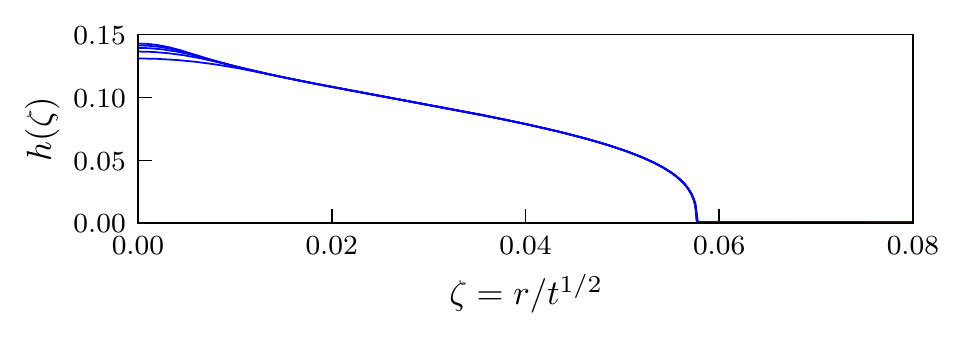}
\end{minipage}%
\begin{minipage}{0.35 \textwidth}
\includegraphics{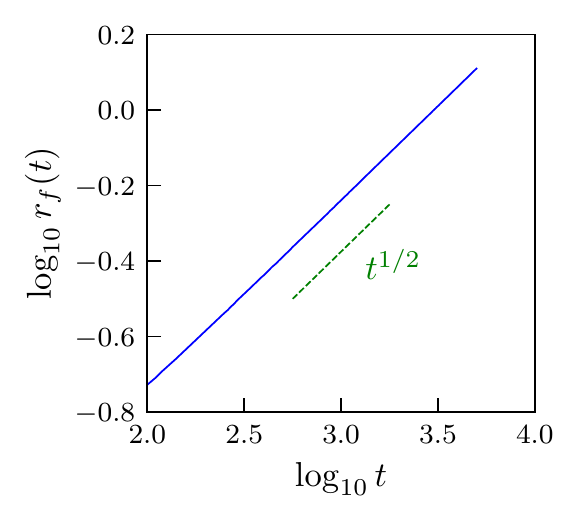}
\end{minipage}
\caption{(left,top) Isoviscous height field evolution in planar geometry.  Snapshots of the flow height profile every 10$^3$ non-dimensional times units are shown.  (left,bottom) Height field evolution scaled by $t^{1/2}$ as a function of the similarity variable $\zeta = r/t^{1/2}$.   Away from the source, solutions converge to the similarity form. (right) Flow front location $r_f$ defined as the non-dimensional height where $h(r_f,t) = 0.01$.  After initial transient behavior \citep{ball19}, numerical results agree with the expected asymptotic scaling.}
\label{fig:plan_bench}
\end{figure}

The source function used for pseudoplasticity (Eq.\ \ref{eq:plan_source}) is adopted for the isoviscous benchmark.  Likewise, the initial and boundary conditions for $h(r,t)$ are the same as those for the pseudoplastic rheology.  In the limit of a constant, point source flux $w_s$, the isoviscous case admits an analytical, similarity solution \citep{huppert82a} in the similarity variable $\zeta = r/t^{1/2}$.  Sufficiently far from the source, numerical solutions converge to the similarity form (Fig.\ \ref{fig:plan_bench}).  The flow front location scales with $t^{1/2}$ as expected from the similarity solution and asymptotic scaling analysis \citep{griffiths93}.    
\begin{figure}[ht!]
\centering
\begin{minipage}{0.65 \textwidth}
\includegraphics{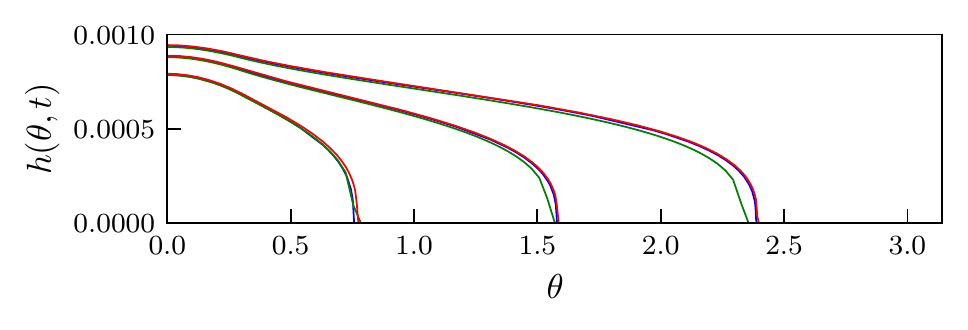}
\end{minipage}%
\begin{minipage}{0.35 \textwidth}
\includegraphics{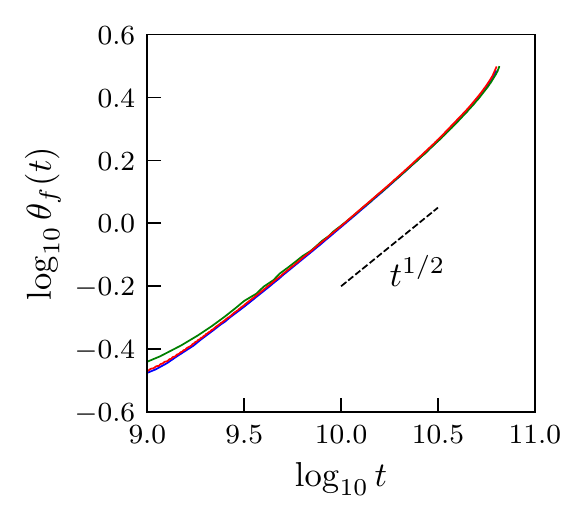}
\end{minipage}
\caption{Comparison of the numerical method for spherical geometry with previous results. (left) Isoviscous height field evolution.  Results are shown for a case with $\theta_w = 0.3$ and non-dimensional source amplitude $w_0^{\prime} = 7.9 \times 10^{-13}$.  Snapshots of the flow height profile for non-dimensional times (0.614, 2.46, 4.77) $\times$ 10$^{10}$. (green) yin-yang grid, (red) adaptive grid, (blue) Python py-pde.  (right) Flow front location $\theta_f(t)$ defined as the angular location where non-dimensional height $h(r_f,t) = 0.01 \; h(0,t)$.  Results are in good agreement throughout spreading. Line colors correspond to the top figure.}
\label{fig:sphr_bench}
\end{figure}    

\subsection{Spherical geometry}

In the absence of a similarity solution in spherical geometry, the numerical method adopted in this work is benchmarked against previous numerical results \citep{reese10,reese11}.  One previous approach uses an adaptive grid scheme designed for nonlinear parabolic equations \citep{blom94} adopted for axisymmetric spreading on a spherical surface.  
Another method utilizes a composite, overlapping  
\citep{chesshire90} ``yin-yang'' grid \citep{kageyama04,kageyama05}.  
This method decomposes the sphere into two component grids, one being a low to mid-latitude portion of a standard spherical-polar grid and the other a rotation of the first and is explicitly two-dimensional in ($\theta$, $\phi$).  That is, in the axisymmetric, isoviscous case (or small strain rate limit for pseudoplastic rheology), the layer height evolution equation Eq.\ (\ref{eq:sphr_evol}) reduces to 
\begin{equation} 
\frac{\partial h}{\partial t} = \frac{1}{12} \nabla^2_{\theta} h^4 + w_s(\theta,t) \; ,
\end{equation} 
where $\nabla_{\theta}$ is the polar angle part of the Laplacian on the unit sphere.
The ``yin-yang'' grid method integrates the evolution equation for flow thickness in time using an explicit Euler scheme and standard, centered, second-order approximations for the full tangential Laplacian operator 
$\nabla^2_{\theta} + \nabla^2_{\phi}$.  The solution at the component grid boundaries are determined by bilinear interpolation from neighboring points.  Source axisymmetry results in axisymmetric layer spreading.           

To accommodate the benchmark, the constant volume flux source function is modified to that used in \citet{reese10}.  In that study, the source function is a truncated gaussian, 
$$
w_s(\theta) = w_0 \; \exp(-\theta^2/\theta_w^2) \; H(\theta_0 - \theta) \; .
$$     
As for the pseudoplastic cases, the initial condition $h(\theta,t=0)$ is a small, finite, smoothly varying function.  The gradient of $h$ at the edges of the domain is set to zero.

Good agreement between methods is found for both 
flow front location as a function of time and flow profile 
as a function of $\theta$.
Also shown is the characteristic scaling (dashed line) for the planar geometry, 
constant volume flux case \citep{huppert82a}.
The flow front eventually converges on the source antipode, $\log \theta_f \approx$ 0.5. 
For sufficiently small distance from the anitpodal axis of symmetry, 
a regime transition occurs \citep{gratton90,diez92} which appears to be resolved in the
solutions (Fig.\ \ref{fig:sphr_bench}).

\end{document}